\documentclass[a4paper]{jpconf}
\usepackage{graphicx}

\def\S{\Sigma}

\def\Aplus{k-\frac{R_+}{n(n-1)} a^2}
\def\Aminus{k-\frac{R_-}{n(n-1)} a^2}
\def\A{k-\frac{R_\pm}{n(n-1)} a^2}

\def\B1{\dot a^2 +\Aplus}
\def\B2{\dot a^2 +\Aminus}
\def\B{\dot a^2 +\A}

\def\be{\begin{equation}}
\def\ee{\end{equation}}
\def\bea{\begin{eqnarray}}
\def\eea{\end{eqnarray}}
\def\bean{\begin{eqnarray*}}
\def\eean{\end{eqnarray*}}

\begin{document}
\title{Double layers in gravity theories}

\author{Jos\'e M M Senovilla}

\address{F\'{\i}sica Te\'orica, Universidad del Pa\'{\i}s Vasco, Apartado 644, 48080 Bilbao, Spain}

\ead{josemm.senovilla@ehu.es}

\begin{abstract}
Gravitational double layers, unlike their classical electromagnetic counterparts, are thought to be forbidden in gravity theories. It has been recently shown, however, that they are feasible in, for instance, gravity theories with a Lagrangian quadratic in the curvature. This is surprising with many potential consequences and the possibility of new physical behaviours. While a clear interpretation seems elusive, several lines of research are open. I present the field equations for double layers, the new physical quantities arising, and several explicit examples.
\end{abstract}

\section{Introduction}
Double layers may be found in several disciplines: 
\begin{enumerate}
\item in biology separating two different forms of matter, 
\item in chemistry as interfaces between different phases (e.g. liquid and solid), 
\item in physics when two laminar parallel shells with opposite electric charges are found next to each other.
\end{enumerate}
They are especially important in plasma and cellular physics, representing abrupt drops in the electric potential by which the cell, or plasma, ``protect'' itself from the environment.
However, there were no known gravitational double layers ---until recently \cite{S,S1}.
Their existence comes as a surprise because an analogy with electrostatics may seem to indicate a dipolar distribution of matter on double layers. At first sight, this sounds strange because only positive masses exist.  
Thus, the potential consequences are tremendous, and new physical behaviours can be described.

\subsection{Reminder: layers and double layers in Classical Electrodynamics}
In classical electromagnetism a surface charge distribution creates an electric field with a jump in its component normal to the surface $S$ \cite{J}. The electrostatic potential $\Phi$ for a volume (with charge density $\rho_V$) plus a surface (with charge density $\sigma$) charge distribution reads:
$$
\Phi (\vec r \, ) = \frac{1}{4\pi \epsilon}\int_V \frac{\rho_V (\vec r\, ')}{|\vec r -\vec r\, '|} dV' + \frac{1}{4\pi \epsilon}\int_S \frac{\sigma (\vec r\, ')}{|\vec r -\vec r\, '|} dA'
$$
Recalling Poisson's equation for the electrostatic potential
$\Delta \Phi =-\frac{\rho}{\epsilon}$
this implies that one can consider the following total charge density
$$
\rho = \rho_V + \sigma \delta_S
$$
where $\delta_S$ is a Dirac delta \underline{with support on $S$}:
$$
\left<\delta_S, f\right> = \int_S f dA
$$
Observe that $\Phi$ is continuous, the electric field $\vec E =-$ grad $\Phi$ has a discontinuity (exclusively in its component normal to the surface) only if $\sigma \neq 0$.

Similarly, the electrostatic potential for a dipole layer charge distribution reads
$$
\Phi (\vec r) = \frac{1}{4\pi \epsilon}\int_S D(\vec r\, ')\,  \vec n \cdot  \vec\nabla '\frac{1}{|\vec r -\vec r\, '|} dA'
$$
where $\vec n$ is the unit outward normal to the layer. Thus, Poisson's equation provides the following charge density
$$
\rho =  \Delta'_S
$$
where $\Delta'_S$ is a `Dirac delta prime' {\it with support on $S$ and strength $D$}:
$$
\left<\Delta'_S, f\right> = - \int_S D\,  \vec n \cdot \vec\nabla f\,  dA
$$
Observe that $\Phi$ is discontinuous now  \cite{J}.

Thus, in electrodynamics, if one wishes to represent objects where the charge or
current is very concentrated the sources are often idealized to charges and currents confined to points, curves or surfaces. A proper mathematical description for such idealized concentrated sources uses the theory of distributions. The sum of distributions and the derivative of a distribution are well defined. 
However, in general the product of two distributions is not well defined. 
This makes the use of distributions
in a nonlinear theory, such as those governing gravity, problematic (Maxwell's equations are linear so that they do not pose any problem).

\section{Spacetimes with distributional curvature}
In gravity theories, one also wishes to have a description of concentrated sources, that is, of concentrated matter and energy. They represent cases with thin shells (or braneworlds, or domain walls) and impulsive waves. However, one cannot simply assume that the metric is a distribution,
since then the curvature (and Einstein) tensor will, in general, not be well defined due to the products of distributions. The solution is to identify the class of metrics whose curvature is well defined as a distribution, and such that the field equations will make sense. For sources on thin shells, an appropriate class of metrics were identified in \cite{I,L,T}. Essentially, these are the metrics which are smooth except on a hypersurface where the metric is only continuous. Although this hypersuface can have any (even changing) causal character \cite{MS}, I will assume herein that it is timelike.

A wider class of metrics was defined in \cite{GT} and called
``regular metrics''. An important result is that if the curvature of a regular metric is concentrated on a submanifold in spacetime, then the submanifold must have co-dimension 1. In other words, regular metrics can represent thin shells but not thin strings, thin textures or thin lines in General Relativity.
The question is, can there be any double layers? In principle they are fine, as they have the right co-dimension. 

To answer this question, let $\Sigma$ be a timelike hypersurface dividing the manifold $V$ into two regions $V^\pm$, and let $n^\mu$ be the unit normal to $\Sigma$ pointing from $V^-$ to $V^+$.
The metrics $g_{\mu\nu}^\pm$ are smooth on $V^\pm$ respectively. Observe that one can actually use two different coordinate systems on $V^\pm$ \cite{MS}, but an indispensable requirement is that the inherited metrics on $\Sigma$ from both sides agree. This implies that there exist local coordinate systems in which the metric is continuous across $\Sigma$ \cite{MS}. Nevertheless, there will be two, in principle different, second fundamental forms of $\Sigma$, one inherited from $V^+$ the other from $V^-$.
Denote them by $K^\pm_{\mu\nu}$ and recall that $n^\mu K^\pm_{\mu\nu}=0$.
Define the discontinuity on $\Sigma$ of the second fundamental form as usual 
 $$
      \left[K_{\mu\nu}\right]\equiv K^{+}_{\mu\nu}-K^{-}_{\mu\nu}, \hspace{1cm} n^{\mu}\left[K_{\mu\nu}\right] =0.
$$
In general, $[P]$ will mean the discontinuity of any object $P$ across $\Sigma$.

Consider the $\Sigma$-step function 
$$
\theta =\left\{
\begin{array}{ccc}
1 &  & V^+\\
1/2 & \mbox{on} & \Sigma \\
0 &  & V^-
\end{array}\right.
$$
Its derivative as a distribution  (distributions are distinguished by an underline) reads 
$$\nabla_\mu\,   \underline{\theta} = n_\mu\,  \underline{\delta}^\Sigma
$$ 
where $\underline{\delta}^\Sigma$ is a scalar distribution (a Dirac delta) with support on $\Sigma$ acting on any test function $Y$ as
$$
\left<\underline{\delta}^\Sigma ,Y\right> =\int_\Sigma Y \, .
$$
The continuity of the metric  allows to compute the Riemann tensor in a distributional sense.
The Riemann tensor acquires, in general, a ``singular" part proportional to $\underline{\delta}^\Sigma$. The explicit expression for the Riemann tensor distribution is \cite{MS}
$$
\underline{R}^\alpha{}_{\beta\mu\nu}=(1-\underline{\theta}) R^{-\alpha}{}_{\beta\mu\nu}+\underline{\theta} R^{+\alpha}{}_{\beta\mu\nu}+\underline{\delta}^\Sigma H^\alpha{}_{\beta\mu\nu} 
$$
where $R^{\pm\alpha}{}_{\beta\mu\nu}$ are the Riemann tensors of $V^\pm$ respectively. 
$H^\alpha{}_{\beta\mu\nu}$ is called the singular part of the Riemann tensor distribution and reads
$$
H_{\alpha\beta\lambda\mu} = - n_{\alpha}\left[K_{\beta\mu}\right]n_{\lambda}+n_{\alpha}\left[K_{\beta\lambda}\right]n_{\mu}-n_{\beta}\left[K_{\alpha\lambda}\right]n_{\mu}+n_{\beta}\left[K_{\alpha\mu}\right]n_{\lambda} \label{HRie}
$$
which is obviously defined only on $\Sigma$. This also leads to the singular part of the Ricci tensor distribution
$$
H^\rho{}_{\beta\rho\mu}\equiv H_{\beta\mu}=-\left[K_{\beta\mu}\right] -\left[K^\rho{}_{\rho}\right] n_\beta n_\mu ,
$$
to the singular part of the scalar curvature
\be
H^\rho{}_{\rho}\equiv H=-2\left[K^\mu{}_{\mu}\right]  \label{H} 
\ee
and to the singular part of the Einstein tensor distribution
$$
{\cal G}_{\beta\mu} = -\left[K_{\beta\mu}\right]+h_{\beta\mu}\left[K^\rho{}_{\rho}\right] , \hspace{1cm} n^\mu {\cal G}_{\beta\mu} =0 \, . 
$$
Here $h_{\beta\mu}=g_{\beta\mu}-n_{\beta}n_{\mu}$ is the projector to $\Sigma$, that is, the space-time version of the first fundamental form on $\S$. The above formulae are purely geometric and therefore they are valid in any theory of gravity. 

In the particular case of General Relativity (GR), the above provides ---via the Einstein field equations--- the singular part $\tau_{\mu\nu}$ of the energy-momentum tensor distribution 
$$
\kappa \tau_{\beta\mu}= -\left[K_{\beta\mu}\right]+h_{\beta\mu}\left[K^\rho{}_{\rho}\right] , \hspace{1cm} n^\mu \tau_{\beta\mu} =0 \quad \mbox{(\underline{only in GR})} 
$$
where $\kappa$ is the gravitational coupling constant.
This is known as the Israel formula. Compare with electrodynamics: here $\tau_{\beta\mu}$ replaces $\sigma$ and the discontinuity of the second fundamental form is the analogue of the discontinuity of the normal component of the electric field.

\subsection{Thin shells in GR}

From the 2nd Bianchi identity (which holds for curvature tensor distributions \cite{MS}) one deduces the formulas 
\bean
(K^+_{\rho\sigma}+K^-_{\rho\sigma}){\cal G}^{\rho\sigma} = 2n^\beta n^\mu \left[ G_{\beta\mu}\right]=2n^\beta n^\mu \left[ R_{\beta\mu}\right]-[R],\\
\overline\nabla^\beta {\cal G}_{\beta\mu}=-n^\rho h^\sigma{}_\mu \left[ G_{\rho\sigma}\right]=-n^\rho h^\sigma{}_\mu \left[ R_{\rho\sigma}\right]  
\eean
where $\overline\nabla$ denotes the intrinsic covariant derivative within $\Sigma$ associated to its first fundamental form. Again, these expressions are geometric and thus valid in general. In GR, {\em but not in general gravity theories}, these can be trivially rewritten in terms of the energy-momentum tensor $T_{\mu\nu}$ and its singular distributional part $\tau_{\mu\nu}$ via the Einstein field equations:
\bean
(K^+_{\rho\sigma}+K^-_{\rho\sigma})\tau^{\rho\sigma} = 2n^\beta n^\mu \left[T_{\beta\mu}\right]\\
\overline\nabla^\beta \tau_{\beta\mu}=-n^\rho h^\sigma{}_\mu \left[T_{\rho\sigma}\right].
\eean
These are the Israel equations \cite{I}.

As one sees explicitly, there is nothing like a delta-prime distribution in GR. Thus, {\em 
gravitational double layers are absent in GR}. This seems very reasonable, because there are only positive masses, gravity is attractive.

\section{Quadratic $F(R)$-gravity}
Bur, what about other gravity theories? One can prove that double layers arise, at least, in any gravity theory with a Lagrangian quadratic on the curvature. In this communication, and in order to illustrate how thin shells behave in alternative gravity theories explicitly showing the feasibility of double layers, I will restrict myself to the simple case of a Lagrangian density given by a quadratic $F(R)$
$$
F(R)=R-2\Lambda +\alpha R^2
$$
where $\Lambda$ (the cosmological constant) and $\alpha$ are constants. GR is the linear case $\alpha =0$. These theories include the important Starobinsky inflationary models \cite{St} and they possess a positive mass theorem \cite{Str}, a well-defined entropy formulation \cite{JKM}, and a well-posed Cauchy problem \cite{TT}. The field equations are \cite{SF,CF,NO}
\be
(1+2\alpha R)R_{\mu\nu}-\frac{1}{2}(R-2\Lambda +\alpha R^2)g_{\mu\nu}
-2\alpha\left( \nabla_\mu \nabla_\nu R -g_{\mu\nu} \nabla_\rho\nabla^\rho R\right)
=\kappa T_{\mu\nu} \label{fe}
\ee
where $\kappa$ is the gravitational coupling constant. Equations (\ref{fe}) make sense as distributions {\em only} if the quadratic terms $R^2$ do not contain singular parts, as one cannot multiply these singular parts. Therefore, these equations demand that $\underline{R}$ cannot have a singular distributional part: $R$ can be, at most, a discontinuous function.

\subsection{Thin shells in quadratic $F(R)$}
Thus,  the field equations make sense only if $H=0$: the singular part of $\underline{R}$ must vanish.
Taking into account formula (\ref{H}) for the singular part $H$ of the scalar curvature this entails inevitably
\be
\left[K^\mu{}_\mu\right] =0 \label{Kcont}
\ee
and implies radical differences with GR \cite{S,S1}: it forbids the use of umbilical hypersurfaces $\S$ ---characterized by $K_{\mu\nu}=f h_{\mu\nu}$--- unless they are totally geodesic. It also restricts the possibility of having mirror symmetric branes to the case with zero mean curvature $K^\mu{}_\mu=0$. 

More importantly, {\em observe that in (\ref{fe}) there are no terms quadratic in $\nabla R$} ! 
This allows for a {\em discontinuous} $R$ and still the field equations make sense.
When $R$ is discontinuous, the terms $\nabla\nabla R$ provide $\underline{\delta}^\Sigma$-terms and derivatives of these: gravitational double layers!

\section{Gravitational double layers}
The energy-momentum tensor distribution can be computed from the field equations (\ref{fe}) leading to 
\be
\underline T_{\mu\nu}=T^+_{\mu\nu} \underline\theta +T^-_{\mu\nu} (1-\underline\theta)+\left(\tau_{\mu\nu}+\tau_\mu n_\nu +\tau_\nu n_\mu +\tau n_\mu n_\nu \right) \underline\delta^{\Sigma} + \underline{t}_{\mu\nu} \label{emt}
\ee
Thus, the singular part of the energy-momentum distribution consists of two different parts, one of thin-shell type proportional to the $\underline\delta^{\Sigma}$, the other of `delta-prime' type, or of dipole layer kind, represented by $\underline{t}_{\mu\nu}$. The former splits into three different objects tangent to $\Sigma$, unlike in GR where only the first term $\tau_{\mu\nu}$ arises.

The explicit expressions and names for the objects in (\ref{emt}) supported on $\S$ are:
\begin{enumerate}
\item the {\em energy-momentum tensor} $\tau_{\mu\nu}$ on $\S$, given by
\be
\kappa \tau_{\mu\nu} =-\{1+\alpha(R^++R^-)\}[K_{\mu\nu}] 
+\alpha \left\{2bh_{\mu\nu}-[R](K^+_{\mu\nu}+ K^-_{\mu\nu})\right\}, \, \, n^\mu\tau_{\mu\nu}=0, \label{tauexc}
\ee
where $b$  is a function on $\S$ measuring the jump on the normal derivative of $R$ across $\S$:
$$
b = n^\mu \left[\nabla_{\mu}R\right] . 
$$
An alternative, useful, expression for (\ref{tauexc}) reads
\be
\kappa \tau_{\mu\nu} =-[K_{\mu\nu}] 
+2\alpha \left(bh_{\mu\nu}-[RK_{\mu\nu}]\right). \label{tauexc1}
\ee
$\tau_{\mu\nu}$ is the only quantity usually defined in standard shells.
\item the {\em external flux momentum} $\tau_{\mu}$ with
\be
\kappa \tau _\mu =-2\alpha \overline\nabla_\mu [R], \quad \quad n^\mu \tau_\mu =0. \label{tauex}
\ee
This momentum vector measures the normal-tangent components of $\underline{T}_{\mu\nu}$ supported on $\S$, so that its timelike component describes the normal flux of energy across $\S$ on $\S$ while its spatial components measure the normal-tangential stresses.
\item the {\em external pressure or tension} $\tau$ defined as
\be 
\kappa \tau = 2\alpha [R] K^\rho{}_\rho , \label{taue} 
\ee
where $K^\rho{}_\rho$ is the trace of {\em either} $K^\pm_{\mu\nu}$ ---they are equal due to (\ref{Kcont}). Taking the trace of (\ref{tauexc1}) one obtains a relation between $b$, $\tau$ and the trace of $\tau_{\mu\nu}$ ($n=$ dim$(\S)$):
$$
\kappa \left(\tau^{\rho}{}_{\rho}+\tau\right) = 2\alpha n b
$$ 
The scalar $\tau$ measures the total normal pressure/tension supported on $\S$.
\item the {\em double-layer energy-momentum tensor distribution} $\underline t_{\mu\nu}$, whose components are defined by acting on any test function $Y$ as follows 
\be
\fbox{$\displaystyle{
\kappa \left<\underline t_{\mu\nu},Y\right> = -2\alpha \int_\Sigma [R]h_{\mu\nu}\,  n^\rho\nabla_\rho Y \, . }$} \label{t}
\ee
\end{enumerate}
This is a symmetric ($\underline t_{\mu\nu}=\underline t_{\nu\mu}$) tensor distribution of Dirac ``delta-prime'' type. $\underline t_{\mu\nu}$ has support on $\S$ but its product with objects intrinsic to $\S$ is not defined unless their extensions off $\S$ are known. For instance, one cannot write $\underline t_{\mu\nu}=h_{\mu\nu} \underline t$ for some scalar distribution $\underline t$ unless $n_\mu$ is extended somehow outside $\S$. This $\underline{t}_{\mu\nu}$ resembles the energy-momentum content of double layer surface charge distributions, or ``dipole distributions'', with strength
$$2\alpha[R]h_{\mu\nu}.
$$
This is the analogue of the strength $D$ in a electrostatic dipole layer.

The case $\alpha=0$ is simply GR supplemented with the condition (\ref{Kcont}). If $\alpha =0$, then the new terms $\tau_{\mu}$, $\tau$ and $\underline{t}_{\mu\nu}$ all vanish and the formulas collapse to those of GR with the extra condition of the continuity of the second fundamental form (\ref{Kcont}).

A term such as (\ref{t}) is remarkable and very surprising. It seems to represent the idealization of an abrupt change on the scalar curvature $R$ which, one has to bear in mind, acts as a source or dynamical variable in quadratic $F(R)$ gravity. Therefore, these gravitational double layers can be thought as an imbalanced strain on the border between two different cosmological constants. 

For the readership who dislikes $F(R)$ theories, it is worth reminding that they are equivalent to a scalar-tensor theory. In particular, quadratic $F(R)$ is equivalent to Brans-Dicke with $\omega =0$ \cite{H,TT,Ch,FT}. The scalar field is $\phi =2\alpha R -1$ with a potential 
$$
V(\phi) = 2\Lambda +\frac{1}{4\alpha} (\phi^2-2\phi -3) .
$$
In this alternative point of view, double layers represent an abrupt discontinuity in $\phi$. Thus, double layers can be thought of as the separation surface of two different phases of $\phi$: a $\phi$-phase transition.

An important remark is that
{\em all the terms} in (\ref{emt}), including $\underline t_{\mu\nu}$, are indispensable to make the entire $\underline T_{\mu\nu}$ divergence free. The field equations derived from this divergence-free property  read:
$$
\overline\nabla^{\nu}\tau_{\mu\nu}+K^{\rho}{}_{\rho}\tau_{\mu}+\overline\nabla_{\mu}\tau =- n^{\nu}h^{\rho}{}_{\mu}[T_{\rho\nu}], $$
\bean
n^{\mu}n^{\nu}[T_{\mu\nu}]-\tau_{\mu\nu}K^{\mu\nu}_{\S}+\overline\nabla^{\mu}\tau_{\mu}
= 2\frac{\alpha}{\kappa} [R]\left(R^{\S}_{\mu\nu}n^{\mu}n^{\nu} +K^{\mu\nu}_{\S}K_{\S\mu\nu} \right) 
\eean
where for any function $f$, $f_\S \equiv (f^+ +f^-)/2$. 
It may be observed that $\tau_{\mu\nu}$ is not divergence-free {\em even} in cases where there is no flux of energy from the bulk across $\S$ (i.e. even when $n^{\nu}h^{\rho}{}_{\mu}[T_{\rho\nu}]=0$) due to the new terms $\tau_\mu$ and $\tau$.  
Notice that $\underline{t}_{\mu\nu}$ is decoupled from the rest of the objects in the field equations.

Two important observations are in order: 
\begin{enumerate}
\item {\em Pure double layers}. 

Given that $\underline{t}_{\mu\nu}$ is does not arise in the field equations,
there seems to be extreme exotic situations in which all $\tau_{\mu\nu}$, $\tau_\mu$ and $\tau$ vanish while $\underline{t}_{\mu\nu}$ does not, describing thin shells with {\em double layer energy-momentum contribution exclusively}. This happens if
$$
K^\rho{}_\rho=0, \hspace{3mm} \overline\nabla_\mu [R]=0, \hspace{3mm} n^\mu \left[\nabla_{\mu}R\right] =0, \hspace{2mm} [K_{\mu\nu}] +2\alpha [RK_{\mu\nu}]=0.
$$
The second and third imply that $R^\pm$ are constants on $\S$, while the first one implies that $\S$ must have zero mean curvature. No examples of these pure double layers have been built hitherto.

\item {\em Matching hypersurfaces in GR are double layers.}

Any thin shell in GR with $[K^\mu{}_\mu]=0$ is a double layer in quadratic $F(R)$. In fact, any properly matched space-time in GR contains a double layer at the matching hypersurface if the true theory is quadratic $F(R)$ ! To see this, recall that in GR a proper matching requires $[K_{\mu\nu}] =0$, but $[R]\neq 0$ in general, and therefore on the matching hypersurface we would have
\bean
\kappa \tau_{\mu\nu} =2\alpha \left(n^\rho \left[\nabla_{\rho}R\right]h_{\mu\nu}-[R]K_{\mu\nu}\right),
\\
\kappa \tau _\mu =-2\alpha \overline\nabla_\mu [R], \hspace{1cm} \kappa \tau = 2\alpha [R] K^\rho{}_\rho ,
\\
\kappa \left<\underline t_{\mu\nu},Y\right> = -2\alpha \int_\Sigma [R]h_{\mu\nu}\,  n^\rho\nabla_\rho Y \, .
\eean
For instance, by matching perfect fluids to vacuum in GR we would have 
$$
[R]= \rho^{GR} |_\S =\mbox{(GR-energy density of the fluid on $\S$)}
$$
This happens because the matching conditions in $F(R)$ theories are more demanding than in GR \cite{DSS,S}.
\end{enumerate}

\section{Examples}
Consider a hypersurface $\S$ separating two constant-curvature $n$-dimensional spacetimes with different cosmological constants $\Lambda^+\neq \Lambda^-$. The line-element on each $\pm$ side reads ((anti)-de Sitter or flat)
$$
ds^2_\pm=-\left(k-\frac{R_\pm x^2_\pm}{n(n-1)} \right)dy^2_\pm+\left(k-\frac{R_\pm x^2_\pm}{n(n-1)} \right)^{-1} dx^2_\pm + x^2_\pm d\Omega^{n-2}_{k} 
$$
where $k=\pm 1,0$ and $d\Omega^{n-2}_{k}$ is the complete Riemannian $(n-2)$-dimensional metric of constant curvature $k$. The cosmological constants $\Lambda_\pm$ are related to the corresponding $\pm$ constant scalar curvatures $R_\pm$ by
$$
\Lambda_\pm =R_\pm \frac{n-2+(n-4) \alpha R_\pm}{2n} .
$$
In general, the ranges (and causal character) of the coordinates $\{x_{\pm},y_{\pm}\}$ depend on the signs of the constants $R_{\pm}$ and $k$. 

To build double layers, one needs to find corresponding hypersurfaces $\S^\pm$ on both $\pm$ sides such that the first fundamental forms agree and the condition (\ref{Kcont})  is satisfied. Restricting to hypersurfaces respecting the spherical symmetry, $\S$ is given by means of the parametric expressions 
$$x_{\pm}=x_\pm(t),Ê\hspace{1cm}  y_{\pm}=y_\pm(t)$$ 
so that the equality of the inherited first fundamental forms on $\S$ requires
$$
x_+(t) = x_-(t)\equiv a(t) $$
and
$$
\frac{dy_\pm}{dt}=\frac{\varepsilon_\pm}{\A}\sqrt{\B}
$$
where $\varepsilon_\pm^2=1$ and dots stand for derivatives with respect to $t$.
Then, the first fundamental form on $\S$ is a Robertson-Walker metric
$$
d\gamma^{2}= -dt^{2}+a^{2}(t) d\Omega^{n-2}_{k}.
$$

The unit normals on each side of $\Sigma$ read
$$
n_\pm =-\epsilon_\pm (-\dot a dy_\pm +\dot y_\pm dx_\pm) 
$$
where the signs $\epsilon_\pm$ determine the part of the $\pm$-side of the bulk space-time  to be matched.
The second fundamental forms on each side read then
\bean
K^\pm_{\mu\nu}dx^\mu dx^\nu = \hspace{5cm} \\
\\
(\varepsilon\epsilon)_\pm \left(-\frac{\ddot a -\frac{R_\pm}{n(n-1)}a}{\sqrt{\B}}\, dt^2 +a\sqrt{\B}\, d\Omega^{2}_{k} \right) .
\eean
The condition $\left[K^\mu{}_\mu\right]=0$  leads to two possible families of solutions for $a(t)$ depending on whether or not $\dot a=0$.
\begin{enumerate}
\item {\bf Case (i):} $a$ is a constant given by the relation
$$
(n-1)^2 (\Aplus)(\Aminus) = 1
$$
with both factors $\A>0$ and $k=-(\epsilon\varepsilon)_+/(\epsilon\varepsilon)_-$. 
\item 
{\bf Case (ii):} $\dot a \neq 0$, then $a(t)$ is the solution of the ODE
\be
\dot a^2 +k = \frac{a^{2n}}{m^2}\left(\frac{[R]}{2n(n-1)}\right)^2+a^2 \frac{R^++R^-}{2n(n-1)}+\frac{m^2}{4a^{2(n-2)}}  \label{equation}
\ee
where $m\neq 0$ is an arbitrary constant. This can also be written in any of the two alternative forms
$$
\B =\left(\frac{m}{2a^{n-2}}\mp\frac{a^n}{2n(n-1)m}[R] \right)^2 . \label{sol}
$$
\end{enumerate}

\subsection{Minkowski/de Sitter double layer}
As a specific example of case (i), consider the possibility that the double layer separates de Sitter from Minkowski space-time, and set the dimension $n=4$. By choosing $R_-=0$, then $k=1$ and 
$$
a^2 =\frac{32}{3} \frac{1}{R_+}= \frac{8}{3}  \frac{1}{\Lambda_+}>0.
$$
This fixes uniquely the constant radius of the double layer in Minkowski, and in de Sitter.
The non-zero eigenvalues of $\tau_{\mu\nu}$ are easily computed to give the energy density and pressure of the double layer (setting $(\epsilon\varepsilon)_+ =-1$)
\bean
\kappa \varrho = 
\frac{8}{3a}(1+2\alpha R_+) ,\\
\kappa p = \frac{1}{3a}
(4+2\alpha R_+)
\eean
which are constant. Concerning the other quantities on the double layer one has  
$$
\tau_\mu =0, 
$$
and
$$
\kappa \tau = 6\frac{\alpha}{a} R_+ =24\frac{\alpha}{a} \Lambda_+= 64 \frac{\alpha}{a^3}.
$$
Finally 
$$
\kappa \left<\underline t_{\mu\nu},Y\right> = -8\alpha\Lambda_+ \int_\Sigma h_{\mu\nu}\,   Y_{,x} 
$$
with a first fundamental form
$$
h_{\mu\nu}dx^\mu dx^\nu = d\gamma^2 = -dt^{2}+\frac{8}{3}  \frac{1}{\Lambda_+} d\Omega^2 \, .
$$
This represents a 3-dimensional Einstein static spacetime.

\subsection{Analysis of the dynamical solution}
In order to analyze the dynamical solution (case (ii)) for a  double layer, one can look at the Eq.(\ref{equation}) in the standard ``kinetic + potential energy" form 
$$\dot a^2 +V(a)=0.
$$ 
In this way it is very simple to get the qualitative behaviour of the solutions $a(t)$. 
The ``potential'' $V(a)$ is an even function. In the physical region with $a>0$, $V(a)$ has a unique maximum and its value is always negative for the cases with $k=0,-1$. 
Therefore, in these cases the double layers describe a $(n-1)$-dimensional Universe starting from a big-bang followed by a decelerated expansion phase until $\dot a$ reaches a minimum value from where the Universe undergoes an accelerated expansion epoch leading to unbounded values of $a$ and $\dot a$.

 For the remaining, closed, case with $k=1$, the maximum of $V(a)$ may be positive or negative depending on whether $|m|$ is small or large, respectively. 
 In the latter case the solutions behave just as in the previous cases with $k=0,-1$. 
 If $|m|$ is small, on the contrary, the possible solutions for $a(t)$ have two branches. 
 In one of them the double layer describes a Universe which, from a big-bang, reaches a re-collapsing time and then contracts to a big crunch. 
The other branch, however, is singularity-free, starting with a large Universe that contracts to a minimum volume and then re-expands with accelerated expansion. 
 Of course, there is a critical value of $|m|$ between the two mentioned possibilities, where the Universe tends asymptotically to an Einstein static solution of the type given in case (i).

Given the properties of the double layer, let us consider that it is a ``braneworld" in a 5-dimensional bulk, that is, set $n=5$. The non-zero eigenvalues of $\tau_{\mu\nu}$ provide the energy density and pressure  within the double layer:
\bean
\kappa \varrho = -(\epsilon\varepsilon)_{+}\left\{\frac{3m}{a^{4}} \left(1+\alpha(R_{+} +R_{-})\right)+\alpha \frac{a^{4}}{4m} [R]^{2} \right\} ,\\
\kappa p = -(\epsilon\varepsilon)_{+}\left\{\frac{m}{a^{4}} \left(1+\alpha(R_{+} +R_{-})\right)-\alpha \frac{a^{4}}{20m} [R]^{2} \right\} 
\eean
Observe that these expressions have two terms, the one on the left dominates for small values of $a$, the other for large values of $a$. 
The former has an equation of state of radiation type $p=\varrho/3$, and is the only one surviving in the GR limit when $\alpha \rightarrow 0$. 
The latter has an equation of state $p=-\varrho/5$ and is proportional to the square of the difference between the constant curvatures at both sides of the double layer. 

Observe that the constant $k$ only is implicit entering only through $a(t)$. Notice also the signs of $R_{\pm}$ are totally free. Thus, any possible combination of de Sitter, anti-de Sitter and flat spacetimes are allowed. Concerning the new energy-momentum quantities we have
\bean
\tau_\mu =0, \hspace{7mm}\\
\\
  \kappa \tau = - (\epsilon\varepsilon)_+
 \frac{2a^{4}}{5m}\alpha [R]^{2}
\eean
and
$$
\kappa \left<\underline t_{\mu\nu},Y\right> = -2\alpha [R] \int_\Sigma h_{\mu\nu}\,  n^\rho\nabla_\rho Y \, .
$$
The last one is independent of the signs $\epsilon_{\pm}$ and $\varepsilon_{\pm}$. 
Here $h_{\mu\nu}$ is the standard Robertson-Walker metric with scale factor $a(t)$.

Observe that  the energy density and pressure satisfy the following {\em generalized continuity equation}
$$
\dot \varrho +3\frac{\dot a}{a} (\varrho +p) =\dot \tau 
$$
which reflects the novel fact that the energy-momentum $\tau_{\mu\nu}$ of the double layer is not divergence free in general. Hence, the traditional behavior of 
$$\varrho\sim a^{-3(1+w)}
$$
 for an equation of state $p=w\varrho$ does not hold for double-layer braneworlds in general.  
This opens the door for new physical behaviors.

As an amusement, one can wonder how advanced physicists living in such a double layer, but unaware of its braneworld/double-layer character, would interpret their measurements, in the same vein as in \cite{MSV,MSV1}. In order to answer this, the energy density and pressure of the double layer must be compared with the GR values as they would be computed by those scientists on $\S$. Imagine, to that end, that they manage to find the correct equation for the scale factor, that is (\ref{equation}) with $n=5$:
$$
\dot a^2 +k = \frac{a^{10}}{m^2}\left(\frac{[R]}{40}\right)^2+a^2 \frac{R_+ +R_-}{40}+\frac{m^2}{4a^6} \, .
$$
This can be rewritten in terms of the Hubble parameter $H(t)$
$$
H^2 +\frac{k}{a^2} = \frac{a^{8}}{m^2}\left(\frac{[R]}{40}\right)^2+ \frac{R_+ +R_-}{40}+\frac{m^2}{4a^8}
$$
so this would just mean that they had been able to determine the behavior of the Hubble parameter with cosmic time $t$. They could then use the Einstein field equations to compute what they would believe are the matter contents of the Universe, and they would get: 
\bean
8\pi G \varrho^{GR} = \frac{3}{4}\left( \frac{a^8}{m^2} \frac{[R]^2}{400}+\frac{m^2}{a^8}\right), \\
8\pi G p^{GR} = \frac{1}{4}\left( \frac{-11a^8}{m^2} \frac{[R]^2}{400}+\frac{5m^2}{a^8}\right),\\
\Lambda^{GR}=\frac{3}{40} \left( R_+ + R_-\right). 
\eean
Now, one can speculate with these formulas...

Note, in any case, that the ``fake''  GR cosmological constant would be simply proportional to the sum of the constant scalar curvatures at both sides of the double layer. Hence, in order to have a positive $ \Lambda^{GR}$, at least one side of the bulk should have a positive constant curvature. Notice also that the choice with one flat Minkowskian side is feasible. And the the value $\Lambda^{GR}=0$ would require a too exact fine tuning between the two sides of the double layer.

\section{Conclusions and prospects}
I have shown that gravitational double layers are consistent in some quadratic theories. If, as usually claimed, the quadratic terms are really needed for a quantum theory of gravity, then double layers may represent an idealized good approximation to situations where the cosmological constant changes abruptly.

Furthermore, good genuine matching hypersurfaces in GR would actually become double layers in any regime where quadratic terms in the Lagrangian start to be non negligible. If, for instance, when approaching a quantum regime these quadratic terms are switched on, or become relevant, then any classical GR solution which can be approximately described by a matching procedure will become an approximate solution which contains, on the matching hypersurface, a gravitational double layer. The implications of this feature must be analyzed in depth. 

Furthermore, the question of how these double layer solutions fit in a general gravitational theory must be addressed, to examine if they actually arise {\em generically} in regimes dominated by quadratic terms in the curvature, or not.

In order to give consistency and a solid background to the gravitational double layers described herein, one should try to find actual, regular, solutions that approach the idealized version described by distributions. Moreover, the stability of the solutions with double layers should also be analyzed. 

Most importantly, a clear interpretation of the dipolar term is needed, and yet unclear.

\section*{Acknowledgements}
The author is grateful to the organizers of the Spanish Relativity Meeting (ERE-2014) for their kind invitation. Supported by grants
FIS2010-15492 (MICINN), GIU12/15 (Gobierno Vasco), P09-FQM-4496 (J. Andaluc\'{\i}a---FEDER) and UFI 11/55 (UPV/EHU).

\end{document}